\documentclass[runningheads]{llncs}
\usepackage{hyperref}
\usepackage{subcaption}
\usepackage{tabularx}
\usepackage{listings}
\usepackage{verbatim}
\usepackage[table]{xcolor}
\usepackage{setspace}
\usepackage{enumitem}
\usepackage{float}
\usepackage[]{caption}
\usepackage{orcidlink}
\usepackage{pdfpages}
\usepackage{enumitem}
\usepackage{graphicx}
\usepackage{booktabs}
\usepackage{amssymb}
\usepackage{rotating}
\usepackage{bbm}
\usepackage{array}
\let\olddefinition\definition
\renewcommand{\definition}{\small\olddefinition}

\title{CPN-Py: A Python-Based Tool \\ for Modeling and Analyzing Colored Petri Nets}
\titlerunning{CPN-Py: A Python-Based Tool for CPNs}

\author{
Alessandro Berti\inst{1}\orcidlink{0000-0002-3279-4795},
Wil M. P. van der Aalst\inst{1}\orcidlink{0000-0002-0955-6940}
}
\authorrunning{A. Berti et al.}

\institute{Process and Data Science (PADS) Chair, RWTH Aachen University, Aachen, Germany \\
\email{\{a.berti,wvdaalst\}@pads.rwth-aachen.de}}
\begin{document}
\maketitle

\begin{abstract}
Colored Petri Nets (CPNs) are an established formalism for modeling processes where tokens carry data. Although tools like CPN Tools and CPN IDE excel at CPN-based simulation, they are often separate from modern data science ecosystems. Meanwhile, Python has become the \emph{de facto} language for process mining, machine learning, and data analytics. In this paper, we introduce \emph{CPN-Py}, a Python library that faithfully preserves the core concepts of Colored Petri Nets---including color sets, timed tokens, guard logic, and hierarchical structures---while providing seamless integration with the Python environment. We discuss its design, highlight its synergy with PM4Py (including stochastic replay, process discovery, and decision mining functionalities), and illustrate how the tool supports state space analysis and hierarchical CPNs. We also outline how \emph{CPN-Py} accommodates large language models, which can generate or refine CPN models through a dedicated JSON-based format.
\end{abstract}

\let\olddefinition\definition
\renewcommand{\definition}{\small\olddefinition}

\section{Introduction}
\label{sec:intro}
Petri nets~\cite{Reisig2013} remain a foundational tool for modeling distributed and concurrent processes. Their intuitive representation of places, transitions, and tokens has driven research on behavioral semantics, verification, and process analysis. Among the many Petri net variants, \emph{Colored Petri Nets (CPNs)}~\cite{Ratzer2003} introduce typed data (color sets), allowing tokens to carry rich information that influences model behavior. This extra expressive power is especially valuable in domains such as distributed systems, communications protocols, and business processes where stateful data is crucial.

Tools like \emph{CPN Tools}~\cite{Ratzer2003} and \emph{CPN IDE}~\cite{Verbeek2021} have long been go-to options for designing, simulating, and analyzing CPNs. However, these platforms often require bridging to modern data analysis ecosystems (e.g., Python-based libraries) for machine learning, process mining, and statistical analysis tasks. Meanwhile, Python has emerged as a de facto standard in these domains, thanks to libraries such as NumPy, pandas, scikit-learn, and \emph{PM4Py}~\cite{Berti2023}, the latter focusing on process mining. 

\begin{figure*}[!t]
\centering
\includegraphics[width=0.9\textwidth]{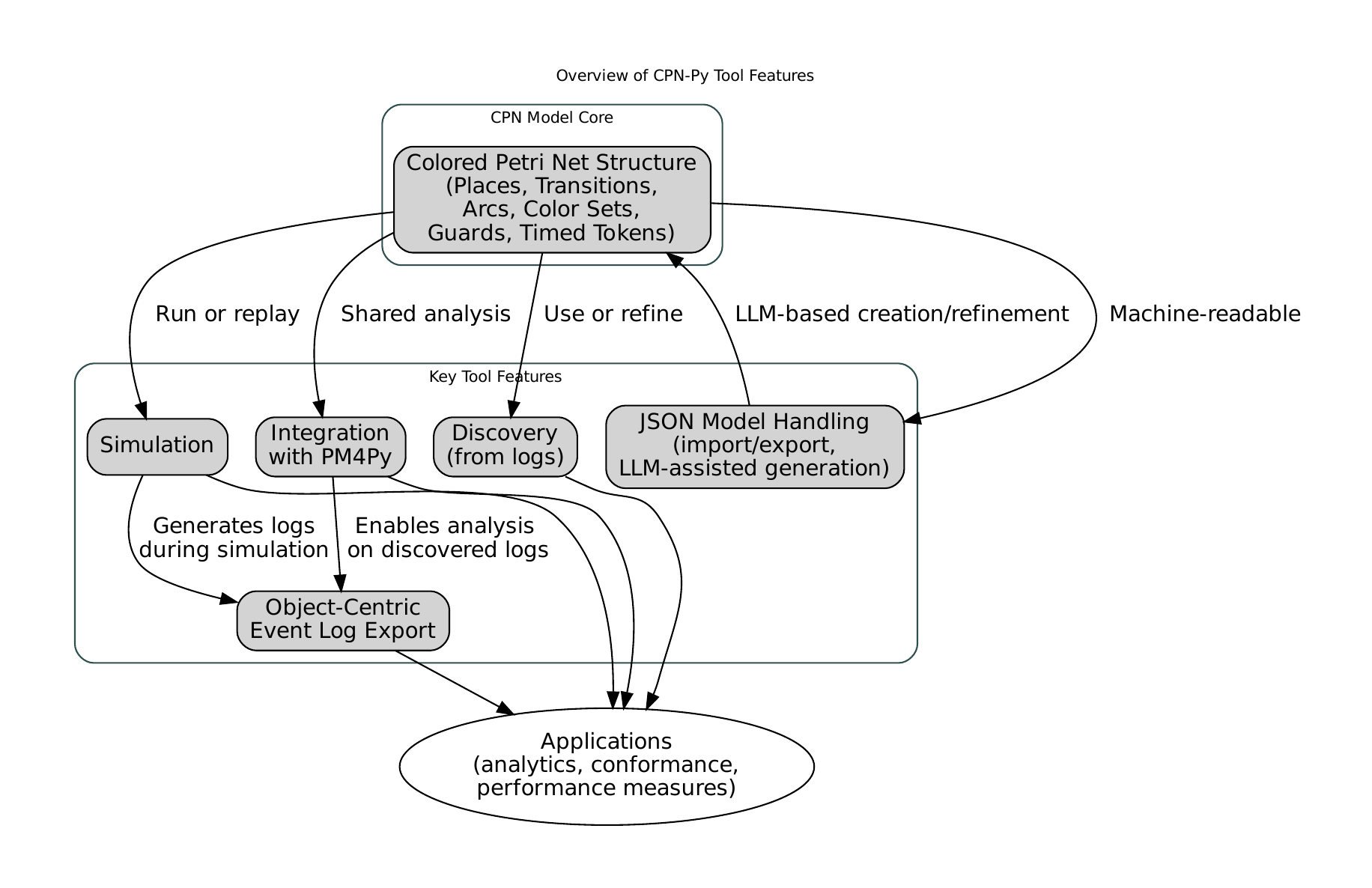}
\caption{High-level view of the tool's features.}
\label{fig:outlinePaper}
\end{figure*}

While some Python libraries provide Petri net implementations, they typically concentrate on either basic Petri net operations or discrete-event simulation~\cite{Dijkman2024}. Their approach can diverge from the \emph{original concept} of Colored Petri Nets, resulting in a simplified or partially formal net model. This gap highlights the need for a Python library that preserves the rigor of classical CPN theory while integrating robustly with data-centric workflows.

\emph{CPN-Py}\footnote{\url{https://github.com/fit-alessandro-berti/cpn-py}} addresses this need by offering a Pythonic approach to constructing, analyzing, and exchanging CPN models. The tool aims to remain consistent with the formal semantics of CPNs, placing strong emphasis on color sets, guard conditions, variable bindings, and timed tokens (when required). Additionally, it offers a dedicated JSON-based interchange format that can be consumed or refined by \emph{large language models (LLMs)} such as GPT-4-like systems. This opens interesting avenues for automated or semi-automated net construction and adaptation. Moreover, the library’s tight coupling with \emph{PM4Py}~\cite{Berti2023} facilitates tasks like process discovery (including decision mining), process mining analysis, stochastic replay, state space analysis, and the generation of \emph{object-centric event logs} (OCEL). Figure \ref{fig:outlinePaper} outlines the features of the tool.

In the following sections, we review related work on Petri net tools and Python-based frameworks, clarify the main motivations for CPN-Py, describe its architecture and capabilities, and illustrate several key applications and research directions (including hierarchical Petri nets, or HCPNs). We conclude by identifying potential opportunities for leveraging LLM-based generation of CPNs, hierarchical net structures, and advanced concurrency optimizations in future extensions.

\section{Related Work}
\label{sec:relwork}

\subsection{From Classical CPN Tools to Python Ecosystems}
\emph{CPN Tools}~\cite{Ratzer2003} remains a reference point for academics and practitioners working with Colored Petri Nets. Its interactive editor and robust simulation engine in Standard ML provide deep support for net construction and state-space analysis. However, extending or integrating these capabilities with Python-based toolchains can require considerable overhead. The shift to \emph{CPN IDE}~\cite{Verbeek2021} introduced a new editor structure in JavaScript and a REST-based architecture around Access/CPN, improving extensibility but retaining a specialized environment largely separate from mainstream Python libraries.

\subsection{Python Libraries for Petri Nets and Simulation}
Python’s popularity in data science has led to multiple libraries offering basic Petri net functionality or process simulation. For instance, \emph{SimPy} provides process-based discrete-event simulation primitives, allowing the modeling of queueing systems or resource constraints. Meanwhile, \emph{SimPN}~\cite{Dijkman2024} is a Python library that explicitly handles timed, colored nets for discrete-event simulation. Although SimPN and certain other libraries demonstrate the viability of Python for CPN-based simulation, the emphasis is usually on discrete-event aspects rather than on the \emph{full} formal structure of CPNs (e.g., enumerated color sets, extensive guard logic, and flexible variable binding). 

\subsection{Process Mining Tools and Interoperability}
\emph{PM4Py}~\cite{Berti2023} has become a core library for process mining. It supports discovery algorithms, conformance checking, log filtering, decision mining, and replay analysis. Nevertheless, the Petri nets typically used in process mining do not fully exploit the power of CPNs (e.g., partial or no color sets). Hence, bridging discovered nets to advanced data-centric modeling requires additional tooling. CPN-Py is designed to interface with PM4Py, supporting the import and export of net structures as well as the production of object-centric event logs (OCEL).

\subsection{LLM-Driven Modeling}
Large language models (LLMs) are increasingly utilized for tasks involving structured data generation and domain-specific modeling~\cite{DBLP:conf/bpm/Berti0A23}. When provided with a well-defined textual or JSON-based schema, modern LLMs can generate or refine process models. Although LLMs have limitations (e.g., potential errors in domain logic, lack of deep semantic understanding), they can provide rapid prototyping assistance or serve as a basis for iterative refinement by human experts. CPN-Py promotes such applications by defining a concise, machine-readable JSON specification for colored Petri nets that could be consumed, validated, or generated in LLM-based workflows.

\section{Motivations}
\label{sec:motivation}

\subsection{Preserving the Formal Essence of CPNs}
A chief motivation for CPN-Py is to maintain the core semantics of Colored Petri Nets within Python. Although existing libraries offer partial solutions, they often diverge from classical definitions: color sets may be replaced by simpler Python datatypes, or guard logic might be restricted. In contrast, CPN-Py incorporates \emph{enumerated}, \emph{product}, and \emph{timed} color sets, along with flexible guard conditions. This ensures that users well-versed in standard CPN theory feel at home, and that teaching or research tasks anchored in CPN theory can be carried out without losing important formal constructs.

\subsection{A Format for Machine-Generated or Machine-Refined Models}
CPN-Py introduces a JSON-based interchange format that fully captures color sets, places, transitions, guard conditions, variable bindings, and initial markings. Rather than focusing solely on human-driven modeling, we consider the potential for large language models (LLMs) to act as co-designers: generating, interpreting, or modifying these JSON files.

While LLMs may not replace human expertise in complex system design, they can accelerate the drafting of process models. For instance, a domain expert can provide textual descriptions, and an LLM may produce an initial CPN in JSON. Later, that model can be validated or refined by humans, ensuring correctness and completeness. This workflow is especially promising when the domain knowledge is scattered across textual documents: the LLM effectively translates partial domain requirements into a structured Petri net model.

\subsection{Less Emphasis on Discrete Event Simulation, More on Integration}
Although CPN-Py offers simulation capabilities (allowing transitions to fire over time, thereby moving tokens and advancing timestamps), the library’s motivations focus on \emph{formal modeling and data-centric integration}. This sets it apart from other Python-based solutions that emphasize discrete-event simulation. The synergy with PM4Py is particularly crucial: once a CPN is enriched with advanced color sets or guard logic, it can generate object-centric logs for conformance analysis or feed into advanced discovery routines. Conversely, discovered nets from PM4Py can be imported as a starting point and elaborated with data semantics (e.g., via decision mining or by adding timing distributions for stochastic replay).

\subsection{Research Accessibility and Ecosystem Benefits}
By remaining entirely in Python, CPN-Py caters to data scientists, process mining researchers, and industrial practitioners who already rely on Python for ETL (Extract-Transform-Load), statistical analysis, or ML tasks. The approach lowers adoption barriers and fosters a single environment for data manipulation, formal modeling, conformance checks, advanced discovery functionalities, state space analysis, and exploration of novel concurrency or data-driven extensions.

\section{Architecture and Features}
\label{sec:arch}

CPN-Py, publicly available at \url{https://github.com/fit-alessandro-berti/cpn-py}, is organized into several interrelated components that collectively provide a Python-native environment for constructing, analyzing, and optionally simulating Colored Petri Nets. This section describes the core design elements, data structures, and utilities that enable a variety of use cases, from formal modeling and process mining integration to JSON-based model generation by large language models. We also highlight \emph{state space analysis} and \emph{hierarchical Petri nets (HCPNs)}.

\subsection{Core Classes and Data Structures}
The foundation of CPN-Py consists of classes that represent the basic entities of a Colored Petri Net:

\begin{itemize}
    \item \textbf{Place:} Each \texttt{Place} is associated with a \emph{color set} (e.g., integer, string, enumerated type). It can hold multiple tokens, each of which must conform to the color set's type. Timed places can also keep track of timestamps when the color set is declared as \emph{timed}.
    \item \textbf{Transition:} A \texttt{Transition} may have:
    \begin{enumerate}
        \item \emph{Variables:} Formal parameters (e.g., \texttt{x, y}) used in guards or arc expressions.
        \item \emph{Guard:} A Python expression that restricts when the transition may fire.
        \item \emph{Transition Delay:} An optional time delay to be added upon firing.
        \item \emph{Arc Expressions:} Python-based expressions for both input (\emph{inArcs}) and output (\emph{outArcs}), which describe how tokens are consumed from and produced to places.
    \end{enumerate}
    \item \textbf{Arc:} Each arc is defined by source and target entities (place-to-transition or transition-to-place), together with an \emph{expression} specifying how tokens move and (optionally) how timestamps are modified.
    \item \textbf{Marking:} A \texttt{Marking} encapsulates the current net state, i.e., the multiset of tokens held by each place at a given time point. Markings are updated whenever a transition fires.
\end{itemize}

\subsubsection{Token Representation and Timestamps}
Tokens in CPN-Py are normally stored as Python objects (e.g., integers, strings, tuples) that adhere to a place’s color set definition. For \emph{timed} color sets, tokens also carry a timestamp. Upon transition firing, the engine can update timestamps according to arc expressions (e.g., using \verb|@+N| to increment time). This mechanism closely follows classic CPN semantics where time can be attached to tokens without requiring a separate global clock object.

\begin{figure*}[!t]
\begin{lstlisting}[language=Python]
from cpnpy.cpn.cpn_imp import CPN, Place, Transition, Arc, Marking, EvaluationContext
from cpnpy.cpn.colorsets import ColorSetParser

# Define color sets
cs_defs = "colset INT = int timed;"
parser = ColorSetParser()
colorsets = parser.parse_definitions(cs_defs)
int_set = colorsets["INT"]

# Create places and a transition
p_in = Place("P_In", int_set)
p_out = Place("P_Out", int_set)
t = Transition("T", guard="x > 0", variables=["x"], transition_delay=1)

# Create arcs: consume 'x' from P_In, produce 'x+1' in P_Out after 2 time units
arc_in = Arc(p_in, t, "x")
arc_out = Arc(t, p_out, "double(x) @+2")

# Build the net
cpn = CPN()
cpn.add_place(p_in)
cpn.add_place(p_out)
cpn.add_transition(t)
cpn.add_arc(arc_in)
cpn.add_arc(arc_out)

# Create a marking
marking = Marking()
marking.set_tokens("P_In", [1, -1])  # both at time 0

# Evaluation context with a user-defined function
user_code = "def double(n): return n*2"
context = EvaluationContext(user_code=user_code)

print("Initial marking:")
print(marking)

# Check enabling
print("Is T enabled?", cpn.is_enabled(t, marking, context))
# True, because x=1 is a positive token.

# Fire the transition
cpn.fire_transition(t, marking, context)
print("After firing T:")
print(marking)
# Token (1) is consumed from P_In, token 2 (double(1)) is added to P_Out with
# timestamp = global_clock + 1 (transition_delay) + 2 (arc delay) = 3

# Advance time
cpn.advance_global_clock(marking)
print("After advancing clock:", marking.global_clock)
# global_clock = 3
\end{lstlisting}
\caption{Example Python code showing a simple CPN-Py usage.}
\label{fig:cpn-python-listing}
\end{figure*}

\subsection{Color Sets and Type Declarations}
Color sets provide the backbone for data handling in a Colored Petri Net. In CPN-Py:
\begin{itemize}
    \item \emph{Base Types:} \verb|int|, \verb|real|, and \verb|string| are considered primitive color sets.
    \item \emph{Enumerated Types:} A developer may define sets of named values (e.g., \{\texttt{red}, \texttt{blue}, \texttt{green}\}) to represent discrete categories.
    \item \emph{Product Types:} Product color sets allow the combination of two previously declared color sets, enabling the representation of composite data (e.g., \verb|(int, string)|).
    \item \emph{Timed Variants:} Appending \verb|timed| to a color set definition implies that all tokens of this type carry a timestamp field.
\end{itemize}

CPN-Py includes methods to parse these definitions directly from Python code or a dedicated JSON structure. Once defined, color sets can be referenced by places, transitions (for guards), or arcs.

\begin{figure*}[!t]
\begin{lstlisting}[language=Python]
from pm4py.objects.log.importer.xes import importer as xes_importer
from cpnpy.discovery.traditional import apply
from cpnpy.cpn.cpn_imp import CPN, Marking, EvaluationContext

# Import an event log using PM4Py
log = xes_importer.apply("my_event_log.xes")

# Run discovery with guard mining enabled
cpn, marking, context = apply(log, parameters={
    "num_simulated_cases": 5,
    "enable_guards_discovery": True
})

print("Constructed CPN:", cpn)
print("Initial Marking:", marking)
print("Evaluation Context:", context)
\end{lstlisting}
\caption{Example Python code showing how to discover a CPN using CPN-Py (including optional decision mining).}
\label{fig:cpn-pro-disc}
\end{figure*}

\subsection{Variable Binding and Guard Logic}
When a transition is checked for \emph{enabling}, the library attempts to bind tokens from input places to the transition’s declared variables. Specifically:
\begin{enumerate}
    \item For each \texttt{inArc}, the engine examines the arc’s expression, along with available tokens, to see if a valid token assignment to variables exists.
    \item If variable assignment is successful, the \emph{guard expression} (if any) is evaluated with these variable bindings. Only if the guard evaluates to \verb|True| does the transition become enabled.
    \item Once enabled, firing the transition consumes the relevant tokens from the input places and produces new tokens at the output places according to the \texttt{outArc} expressions. Any time increment specified in an \verb|@+N| annotation is applied at this point.
\end{enumerate}

To permit rich data-dependent behavior, guard expressions can invoke user-defined functions or refer to constants and data structures placed in an \\
\texttt{EvaluationContext}. This context can be loaded from a Python file or passed programmatically, enabling domain-specific logic (e.g., calling external libraries).

\subsection{Discovery from Event Logs, Stochastic Replay, and Decision Mining}
CPN-Py provides a dedicated functionality \\ (\texttt{cpnpy.discovery.traditional.apply}) that interfaces with PM4Py to discover a CPN model from a classical event log, following an approach similar to \cite{DBLP:journals/sttt/RozinatMSA08}. This feature:
\begin{enumerate}
    \item Uses a \emph{process discovery} algorithm from PM4Py to obtain a Petri net (applying the inductive miner by default).
    \item Offers \emph{decision mining} capabilities, generating guard expressions for transitions based on data attributes extracted from the log.
    \item Supports the configuration of an initial marking based on real cases or synthetic data, enabling \emph{stochastic replay} if desired.
\end{enumerate}
Figure~\ref{fig:cpn-pro-disc} shows sample usage. The returned \texttt{EvaluationContext} can handle custom Python functions for guard evaluation, as well as stochastic distributions for timed behaviors.

\begin{figure*}[!t]
\footnotesize
\begin{lstlisting}[language=Python]
from cpnpy.analysis.analyzer import StateSpaceAnalyzer
from cpnpy.cpn.cpn_imp import CPN, Marking, EvaluationContext

# Define a CPN, marking, and context
cpn = CPN()
# ... add places, transitions, arcs ...

marking = Marking()
# ... set initial tokens ...

context = EvaluationContext()

# Build the analyzer
analyzer = StateSpaceAnalyzer(cpn, marking, context)

# Compute and retrieve summary statistics
report = analyzer.summarize()
print("=== State Space Report ===")
for key, val in report.items():
    print(f"{key}: {val}")
\end{lstlisting}
\caption{Example usage of the built-in \texttt{StateSpaceAnalyzer} for reachability and SCC analysis.}
\label{fig:ssa-code}
\end{figure*}

\subsection{Simulation and Execution Model}
Although CPN-Py is not centered purely on discrete-event simulation, it supports step-by-step or automated firing sequences:
\begin{itemize}
    \item \emph{Manual Step:} The user can query all \emph{enabled transitions}, select one (or more) transitions to fire, and update the marking accordingly. 
    \item \emph{Time Progression:} If timed tokens are used, each transition may advance the net’s \emph{simulation clock} or add a delay upon firing. Successive firings occur in chronological order if multiple events exist at different timestamps.
    \item \emph{Extended Logging:} During simulation, event logs that capture each firing can be generated, optionally referencing the tokens consumed and produced. This feature is fundamental for subsequent analysis in PM4Py or other process mining tools.
\end{itemize}

This flexibility accommodates various modeling scenarios, from immediate transitions with no time semantics to data-driven workflows where transitions may be delayed or prevented by complex guard logic.

\subsection{State Space Analysis with the \texttt{StateSpaceAnalyzer}}
Beyond discovery and simulation, CPN-Py offers a built-in \texttt{StateSpaceAnalyzer} that can build the \emph{reachability graph (RG)} and the \emph{strongly connected components (SCC)} graph of a CPN. It supports:
\begin{itemize}
    \item Identification of \emph{home markings} (markings appearing in every infinite firing sequence).
    \item Detection of \emph{dead markings} (markings with no enabled transitions).
    \item Determination of \emph{live}, \emph{dead}, and \emph{impartial} transitions (based on terminal SCC analysis).
    \item \emph{Place bounds} extraction (minimum and maximum number of tokens) and other structural or behavioral properties.
\end{itemize}
These capabilities provide insights into liveness, boundedness, and other verification properties that are central in Petri net theory.

\subsection{Hierarchical Petri Nets (HCPNs)}
A distinguishing feature of advanced CPN formalisms is the support for \emph{hierarchical modeling}. In CPN-Py, \emph{hierarchical nets} (HCPNs) introduce:
\begin{itemize}
    \item \emph{Substitution Transitions}: Special transitions that delegate token flow to another \emph{submodule} or child CPN, allowing multi-level modular designs.
    \item \emph{Fusion Sets}: Mechanisms to fuse places across modules for shared state or data. 
    \item \emph{Visualization Tools}: Graphviz-based rendering that can show parent and child modules, with dashed edges linking substitution transitions to the submodules.
\end{itemize}
This hierarchical approach enables more scalable and maintainable models, particularly for complex processes with repeating sub-process structures.

\subsection{Interoperability with CPN Tools (CPN XML)}
\label{subsec:cpnxml}

Although CPN-Py internally uses a JSON-based format for storing Colored Petri Nets, it also supports \emph{importing} and \emph{exporting} CPN Tools' XML files (\emph{CPN XML}). This ensures basic compatibility with the official CPN Tools format~\cite{Ratzer2003} and allows both legacy and newly created models to be transferred between the two environments.

\subsubsection{From CPN XML to JSON}

When converting from CPN Tools' XML representation to CPN-Py's JSON format, the library:

\begin{enumerate}[leftmargin=2em]
\item \emph{Maps structural elements} (places, transitions, arcs) directly into the JSON schema.
\item \emph{Translates Standard ML} expressions (guards, arc expressions) into Python. Since CPN Tools uses Standard ML for expressions, CPN-Py offers utility functions to \emph{automatically attempt} an SML-to-Python conversion by leveraging large language models \\ (via the module \verb|cpnpy.util.conversion.llm_json_fixing|). 
\item \emph{Generates a complete JSON definition} that can be loaded by CPN-Py for subsequent analysis, simulation, or refinement.
\end{enumerate}

A reference script, \verb|importing_mynet.py|, illustrates end-to-end usage (\emph{see} \texttt{examples/conversion/xml\_to\_json/importing\_mynet.py}). The script is shown in Listing~\ref{lst:xmltojson}.

\begin{figure}[!t]
\begin{lstlisting}[language=Python]
from cpnpy.util.conversion import cpn_xml_to_json
from cpnpy.cpn import importer
from cpnpy.visualization import visualizer
import json

json_path = "xml_to_json.json"

if __name__ == "__main__":
    dct = cpn_xml_to_json.cpn_xml_to_json("files/other/xml/mynet.cpn")
    json.dump(dct, open(json_path, "w"))

    dct = json.load(open(json_path, "r"))

    cpn, marking, context = importer.import_cpn_from_json(dct)
    print(cpn)
    print(marking)
    viz = visualizer.CPNGraphViz()
    viz.apply(cpn, marking, format="svg")
    viz.view()
\end{lstlisting}
\caption{Importing a CPN Tools XML into CPN-Py JSON.}
\label{lst:xmltojson}
\end{figure}

\subsubsection{From JSON to CPN XML (Stub)}

Conversely, CPN-Py can generate a \emph{minimal} CPN Tools XML file from a JSON-based net description. This creates a ``stub'' XML that typically needs some manual modifications to leverage full \emph{CPN Tools}-specific features (e.g., graphical layout, layout annotations, or advanced color set declarations incompatible with the JSON schema). Listing~\ref{lst:jsontoxml} shows an illustrative snippet (an expanded example is available in \texttt{examples/conversion/json\_to\_xml/auto\_discovery.py}).

\begin{figure}[!t]
\begin{lstlisting}[language=Python]
import pm4py
from cpnpy.discovery import traditional as traditional_discovery
from cpnpy.cpn import exporter
from cpnpy.util.conversion import json_to_cpn_xml

json_path = "auto_disc1.json"
xml_path = "auto_disc1.cpn"

if __name__ == "__main__":
    log = pm4py.read_xes("files/other/xes/running-example.xes")
    cpn, marking, context = traditional_discovery.apply(log, parameters={"enable_guards_discovery": False, "enable_timing_discovery": False})
    exporter.export_cpn_to_json(cpn, marking, context, json_path)
    json_to_cpn_xml.apply(json_path)

    xml = json_to_cpn_xml.apply(json_path)

    F = open(xml_path, "w")
    F.write(xml)
    F.close()
\end{lstlisting}
\caption{Exporting a CPN-Py JSON definition to a stub CPN Tools XML.}
\label{lst:jsontoxml}
\end{figure}

In practice, once the stub XML is generated, you can open it in CPN Tools (or \emph{CPN IDE}) for further visual refinements or advanced semantic edits that go beyond CPN-Py’s scope. This two-way interoperability---importing refined CPN Tools models and generating XML stubs from CPN-Py---helps practitioners preserve their existing workflows while still benefiting from Python-based simulation, data mining, or AI-driven techniques in CPN-Py.

\subsection{Streamlit-Based Graphical Interface}
\label{subsec:gui}

CPN-Py offers a prototypal, \emph{Streamlit}-based web interface that allows users to \emph{interactively} create, visualize, and simulate Colored Petri Nets. Through a browser view, you can perform typical modeling tasks without writing Python code directly, including:

\begin{figure*}[!t] 
    \begin{minipage}{0.48\textwidth} 
        \centering
        \includegraphics[width=\textwidth]{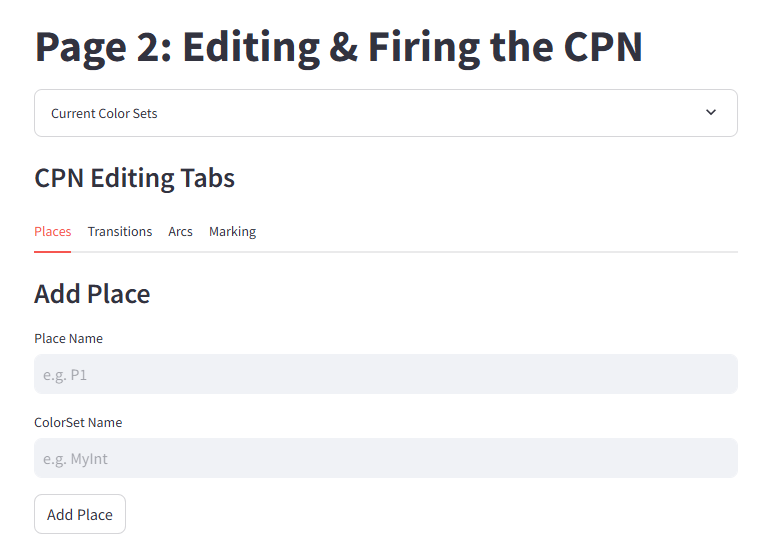} 
        \caption{The prototypal interface allows inserting new places, transitions, and arcs.}
        \label{fig:cpnAdding}
    \end{minipage}
    \hfill 
    \begin{minipage}{0.48\textwidth}
        \centering
        \includegraphics[width=\textwidth]{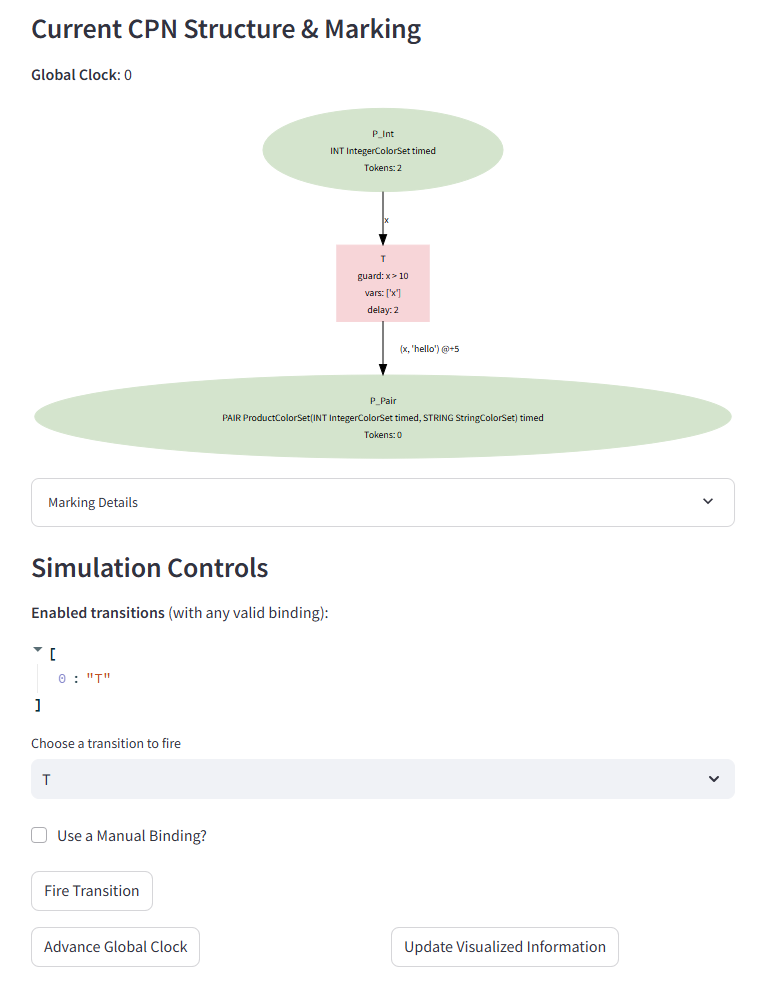}
        \caption{The prototype interface visualizes the CPN structure and marking. Simulation controls are also provided.}
        \label{fig:cpnViewMarking}
    \end{minipage}
\end{figure*}

\begin{itemize}
  \item \emph{Importing} an existing CPN from a JSON file.
  \item \emph{Defining} new color sets from scratch.
  \item \emph{Adding} places, transitions, arcs, and initial tokens.
  \item \emph{Firing} transitions or \emph{advancing} the global clock step-by-step.
  \item \emph{Visualizing} the net and its current marking (tokens, global clock) in a Graphviz diagram.
  \item \emph{Exporting} the resulting CPN to a JSON file for future reuse or processing.
\end{itemize}

\subsubsection{How to Start the Interface}
\label{subsubsec:run-gui}

To launch the Streamlit interface:
\begin{enumerate}
  \item \emph{Navigate} to the root folder of your project (the directory containing the \texttt{cpnpy/} subfolder).
  \item \emph{Run} the command (for Windows users) in a terminal: \\ \texttt{streamlit run ./cpnpy/home.py}
  \item \emph{Open} the provided URL in a web browser. The interface will display separate pages to:
  \begin{itemize}
    \item \emph{Import} a JSON-based CPN or \emph{Create} color sets from scratch (Page 1).
    \item \emph{Edit} the net by adding places, transitions, arcs, and marking information (Page 2, see Figure \ref{fig:cpnAdding}).
    \item \emph{Simulate} the net by firing transitions or advancing the clock (see Figure \ref{fig:cpnViewMarking}).
    \item \emph{Visualize} the current marking via an automatically updated Graphviz diagram (see Figure \ref{fig:cpnViewMarking}).
    \item \emph{Export} the modified net to JSON.
  \end{itemize}
\end{enumerate}

Although this GUI is still in a prototype phase, it demonstrates how CPN-Py can be integrated into an interactive environment, bridging the gap between code-driven modeling and a more accessible, point-and-click experience for building and exploring Colored Petri Nets.

\section{Use Cases and Extensions}
\label{sec:usecases}

\subsection{Integrating Data Semantics into Discovered Nets}
Process discovery from real-life event logs typically yields a Petri net that captures control-flow but not detailed data logic. Researchers or practitioners can import this net into CPN-Py, specify color sets for product types or user roles, add guard conditions reflecting business rules, and then simulate or analyze the enriched model. This process-based extension provides deeper insights into how specific data conditions drive particular transition firings. Stochastic replay can further model varied timing distributions or branching probabilities.

\subsection{LLM-Assisted Drafting of Models}
A domain expert might describe a workflow in natural language (e.g., “Parts heavier than 50 grams require a quality recheck, after which they are sorted by color”). A large language model can transform such statements into a JSON-based CPN specification, naming the color sets (\emph{WeightSet} for numerical data, \emph{ColorEnum} for color categories), setting up places (e.g., \emph{WeighingStation}, \emph{QualityControl}), and transitions with corresponding guard expressions. The resulting JSON can be validated against the schema, loaded into CPN-Py, and then edited by an expert. While not foolproof, this approach can accelerate early-stage modeling or help novices produce syntactically valid CPN definitions.

\subsection{Instructional Use and Rapid Prototyping}
Courses on Petri nets or process mining often require hands-on exercises. Providing students with Python notebooks that integrate CPN-Py alongside PM4Py can streamline teaching. Students can create, export, and analyze small colored nets, perform conformance checks on generated logs, or experiment with partial expansions of discovered nets (e.g., adding hierarchical submodules).

\subsection{Future Research Directions}
Beyond the initial scope, CPN-Py lays the groundwork for more advanced functionalities:
\begin{itemize}
    \item \emph{Extended Hierarchical Net Support}: Substitution transitions referencing entire sub-nets would help model large systems and advanced multi-level processes.
    \item \emph{Parallel Simulation Optimizations}: Researchers interested in performance could build concurrency or distributed simulations on top of CPN-Py. 
    \item \emph{Advanced Verification Methods}: Symbolic or partial-order-based state-space analysis may be explored within Python, benefiting from existing libraries like NetworkX for graph analysis.
    \item \emph{Deeper LLM Integration}: With refined prompting strategies, LLMs may handle not just net generation, but also net adaptation, conflict resolution, or insertion of concurrency constructs. Human experts would still guide the final design, but the synergy could be powerful.
\end{itemize}

\section{Conclusion}
\label{sec:conclusion}
In this paper, we introduced \emph{CPN-Py}, a Python library devoted to reflecting the original concepts of Colored Petri Nets in a data-centric environment. By preserving formal structures---color sets, guard logic, timed tokens, hierarchical transitions---while supporting a new JSON-based model format, CPN-Py positions itself for easy integration with Python’s rich ecosystem. Emphasis lies in bridging formal modeling and process mining, rather than focusing solely on discrete-event simulation. Researchers and practitioners can rely on CPN-Py to create, adapt, and share net definitions, taking advantage of large language models for draft generation or refinement. PM4Py integration enables discovery, conformance checking, stochastic replay, decision mining, and object-centric log generation, while the built-in \texttt{StateSpaceAnalyzer} provides insight into net properties and behavioral correctness.

Future work includes more extensive support for hierarchical nets, improved concurrency management, and deeper expansions of the LLM-driven design cycle. We envision CPN-Py as an evolving resource, aligning rigorous theoretical foundations with modern data analytics, process mining, and AI-driven modeling, thereby stimulating further innovation in both academic and industrial settings.


\bibliographystyle{splncs04}
\bibliography{references}

\begin{thebibliography}{1}
\providecommand{\url}[1]{\texttt{#1}}
\providecommand{\urlprefix}{URL }
\providecommand{\doi}[1]{https://doi.org/#1}

\bibitem{DBLP:conf/bpm/Berti0A23}
Berti, A., Schuster, D., van~der Aalst, W.M.P.: {Abstractions, Scenarios, and
  Prompt Definitions for Process Mining with LLMs: A Case Study}. In: Weerdt,
  J.D., Pufahl, L. (eds.) Business Process Management Workshops - {BPM} 2023
  International Workshops, Utrecht, The Netherlands, September 11-15, 2023,
  Revised Selected Papers. Lecture Notes in Business Information Processing,
  vol.~492, pp. 427--439. Springer (2023),
  \url{https://doi.org/10.1007/978-3-031-50974-2\_32}

\bibitem{Berti2023}
Berti, A., van Zelst, S.J., Schuster, D.: {PM4Py: A process mining library for
  Python}. Softw. Impacts  \textbf{17},  100556 (2023),
  \url{https://doi.org/10.1016/j.simpa.2023.100556}

\bibitem{Dijkman2024}
Dijkman, R.M.: {SimPN: A Python Library for Modeling and Simulating Timed,
  Colored Petri Nets}. In: del{-}R{\'{\i}}o{-}Ortega, A., Montali, M.,
  Rinderle{-}Ma, S., Reijers, H.A., vom Brocke, J., Weske, M., Depaire, B.,
  Indulska, M., van~der Aa, H., Adrian, W.T., Genga, L., Leemans, S.J.J.,
  Gdowska, K., G{\'{o}}mez{-}L{\'{o}}pez, M.T., Rehse, J., Agostinelli, S.
  (eds.) Proceedings of the Best Dissertation Award, Doctoral Consortium, and
  Demonstration {\&} Resources Forum at {BPM} 2024 co-located with 22nd
  International Conference on Business Process Management {(BPM} 2024), Krakow,
  Poland, September 1st to 6th, 2024. {CEUR} Workshop Proceedings, vol.~3758,
  pp. 71--75. CEUR-WS.org (2024),
  \url{https://ceur-ws.org/Vol-3758/paper-12.pdf}

\bibitem{Ratzer2003}
Ratzer, A.V., Wells, L., Lassen, H.M., Laursen, M., Qvortrup, J.F., Stissing,
  M.S., Westergaard, M., Christensen, S., Jensen, K.: {CPN Tools for Editing,
  Simulating, and Analysing Coloured Petri Nets}. In: van~der Aalst, W.M.P.,
  Best, E. (eds.) Applications and Theory of Petri Nets 2003, 24th
  International Conference, {ICATPN} 2003, Eindhoven, The Netherlands, June
  23-27, 2003, Proceedings. Lecture Notes in Computer Science, vol.~2679, pp.
  450--462. Springer (2003), \url{https://doi.org/10.1007/3-540-44919-1\_28}

\bibitem{Reisig2013}
Reisig, W.: {Understanding Petri Nets - Modeling Techniques, Analysis Methods,
  Case Studies}. Springer (2013),
  \url{https://doi.org/10.1007/978-3-642-33278-4}

\bibitem{DBLP:journals/sttt/RozinatMSA08}
Rozinat, A., Mans, R.S., Song, M., van~der Aalst, W.M.P.: Discovering colored
  petri nets from event logs. Int. J. Softw. Tools Technol. Transf.
  \textbf{10}(1),  57--74 (2008),
  \url{https://doi.org/10.1007/s10009-007-0051-0}

\bibitem{Verbeek2021}
Verbeek, E., Fahland, D.: Cpn ide: An extensible replacement for cpn tools that
  uses access/cpn. In: 3rd International Conference on Process Mining, ICPM
  2021. pp. 29--30. CEUR-WS. org (2021),
  \url{https://research.tue.nl/files/201243951/demo_197.pdf}

\end{thebibliography}

\end{document}